# Negative Absolute Temperatures

Quanmin Guo

*School of Physics and Astronomy, University of Birmingham, Birmingham B15 2TT, UK*

**Abstract**

The concept of negative absolute temperature, introduced by Ramsey based on the study of a nuclear spin system by Purcell and Pound in 1951, has been subject to continued debate. According to a recent analysis by Struchtrup, the apparent negative temperature states are temperature unstable states for which no temperature-positive or negative-can be defined. Ramsey was aware of the potential problem with his treatment of the negative temperature by pointing it out that the apparent negative temperature states cannot be achieved via any reversible process. In this paper, we demonstrate that the existence of an upper limit in the allowed states of the system is not a sufficient condition, albeit a necessary condition, for the appearance of negative temperatures. A thermal system cannot move into any negative temperature state via reversible thermodynamic processes. Negative temperature of a system is a direct outcome of incorrectly applying thermodynamic functions to an irreversible process.

The entropy change of a system is described by $dS = \dfrac{dQ}{T}$, where $dQ^1$ and $T$ are the heat added to the system and the temperature of the system, respectively. Incorporating the first law of thermodynamics, $dU = dQ - dW$, we get:

$$dS = \frac{dU + dW}{T} \quad (1)$$

and

$$\frac{1}{T} = \frac{dS}{dU + dW} \quad (2)$$

In the above equations, $dU$ is the change to the internal energy of the system, and $dW$ the amount of work performed by the system. For a reversible process involving no work, Equation (2) becomes:

$$\frac{1}{T} = \left(\frac{\partial S}{\partial U}\right)_{dw=0} \quad (3)$$

For a fluidic system:

$$\frac{1}{T} = \left(\frac{\partial S}{\partial U}\right)_{N,V} \quad (4)$$

Keeping $V$ and $N$ constant in Eq. (4) ensures that the partial derivative, $\left(\dfrac{\partial S}{\partial U}\right)_{N,V}$, has no contribution from work or particle number changes.

For a system with an upper limit in its allowed states, $\left(\dfrac{\partial S}{\partial U}\right)_{N,V}$ can turn into negative leading to the introduction of negative absolute temperatures for spin systems [1-3] and quantum gases [4]. In a recent paper [5], Struchtrup critically analyzed systems in inverted quantum states and described them as temperature unstable states, for which no temperature can be defined. The work of Struchtrup represents the most recent challenge to the validity of negative temperature which has been the point of a continued debate for over half a century [6-11]. One of the conditions for a system to exhibit apparent negative temperature is that its energy structure must have an upper bound. The spin system with just two discrete energy levels is the first and only example of such a system for many years until the recent report of negative temperature in ultra-cold gases [4]. Introducing the negative temperature to the spin system, Ramsey [2] considered entropy $S$ as a function of $U$ and found that $\left(\frac{\partial S}{\partial U}\right)$ changes from positive to negative as temperature "passes" infinity. The essential requirements for a system to be capable of negative temperature are given by Ramsey [3]: i) The elements of the thermodynamical system must be in thermodynamical equilibrium among themselves in order that the system can be described by a temperature at all; ii) there must be an upper limit to the possible energy of the allowed states of the system; and iii) the system must be thermally isolated from all systems which do not satisfy both the above conditions. The three requirements above can be regarded as the necessary conditions for negative temperature. However, they are not the sufficient conditions. For example, in order to make a spin system into a negative temperature state, one has to follow a special procedure that involves a sudden reversal of the external magnetic field [1]. The net magnetization of the system remains unresponsive to such a rapid change of field. This puts the system into a super energetic state which is hotter than $T = \infty$. The negative temperature states of the spin system cannot be reached via any reversible processes, as pointed out by Ramsey. This presents a major problem in the interpretation of the negative temperature states. In thermodynamics, any equilibrium state must be accessible via a reversible process. The use of equations such as Eq. (3) and (4) to evaluate the temperature of the system is based on the assumption that the state is accessible via reversible processes. Some recent debates about the nature of negative temperatures are around the different definitions of entropy [7]. We argue here that because the population-inverted states cannot be reached via any reversible process, it is not appropriate to apply Eq. (4) to get the temperature. It is the incorrect application of thermodynamic functions to an irreversible process that leads to the appearance of negative temperatures. While the negative temperature concept is useful to analyze how the entropy of a spin system changes with magnetic energy, no physical system can exist in a negative temperature state.

---

[1] We use dQ and dW for the infinitesimal changes in heat and work, respectively, with the understanding that neither heat nor work is function of states.

The change of the magnetic energy for a spin system is:

$$dU = TdS - MdB \qquad (5)$$

where M magnetic moment of the system, This leads to the spin temperature:

$$\frac{1}{T} = \left(\frac{\partial S}{\partial U}\right)_B \qquad (6)$$

If the magnetic field is zero, the spin system is unable to take or release energy. Therefore, the spin temperature is not defined for zero field. For a finite $B$ field, there is a finite $M$ corresponding to a specific temperature. As $T$ increases, $M$ decreases and the entropy of the spin system increases. When $T$ approaches infinity, $M$ is zero and the entropy of the spin system is maximized. Consider the system under equilibrium at infinite temperature, it is in equilibrium with a background of black body radiation. The energy density of the blackbody radiation $\rho(f) = \infty$ for all frequencies of $f$ corresponding to infinite $T$. The spin system would move into the negative temperature state if it could absorb more energy from the surrounding. However, this cannot be done. One cannot make the surrounding to have an energy density higher than infinity. This puts an upper limit for the temperature of any system to be infinity and no system can be hotter than this. One may argue that the spin system can still absorb energy from a reservoir which is at negative $T$ and hotter than the system. However, such a negative $T$ reservoir must have energy density $\rho(f) > \infty$. This is physically unrealistic though. The interaction of the spin system with a background of black body radiation has not been discussed much since it was mentioned in the paper of Ramsey. Since the apparent negative temperature state of a spin system is only achieved using the special procedure of rapid reversal of an external magnetic field, we will discuss in the following what are the physical consequences of rapid field reversal.

We first consider spins completely decoupled from other systems. The spins are assumed to interact only with the applied magnetic field. When a magnetic field, $B$, is applied to $N$ non-interacting spins, full alignment of spins is expected such that the total magnetic moment of the system is $M = N\mu$ where $\mu$ is the magnetic moment associated with each spin. The magnetic energy of the system is $U = -BN\mu$. The energy of the system depends on the magnitude of the $B$ field, but the entropy is zero and independent of $B$ as long as $B$ is non zero. Each spin has $-\mu B$ energy which cannot be exchanged with other spins. Such a system can be sufficiently described by its inertial energy $U = -BN\mu$ without the need to know its temperature, although we could assign $T = 0K$ to the system because of the complete alignment of the magnetic moments. If such a system is brought into thermal contact with a reservoir at 0 K, there will be no net energy flow between them. If we reverse the $B$ field, the magnetic moments should flip and align to the direction of the new field. The system behaves the same regardless the direction of the $B$ field, with the only difference being the direction of the magnetic moment.

We now examine the conditions used to prepare population inverted spin systems. The spin system is initially under thermal equilibrium at $T$ with the surrounding which can be regarded as the lattice. With an external $B$ field, the magnetic moment of the system is $M = N\mu \tanh(\frac{\mu B}{kT})$ according to the Boltzmann distribution. If the $B$ field is reduced reversibly to zero and then inverted while the spin system is in contact with the surrounding, then $M$ would be reduced gradually to zero and then change direction. During the above process, the spin system remains at constant $T$ while its magnetic moment changes from $M = N\mu \tanh(\frac{\mu B}{kT})$ to $M = -N\mu \tanh(\frac{\mu B}{kT})$. At any moment, the state of the system can be quantified by (B,T) and there is no need in this case for the introduction of a negative $T$. The apparent negative temperature states were achieved by rapidly reversing the magnetic field from an equilibrium state (B,T) [1]. The magnetic moment of the system remains unchanged during field reversal. A question to ask is what is holding the magnetic moment unresponsive to the changing $B$ field? The standard interpretation [2] is that the spin system is effectively isolated from the lattice because of a very long spin-lattice relaxation time. This interpretation is not consistent with our earlier analysis for an isolated spin system. If the spins are completely isolated from the surrounding apart from being allowed to interact with the $B$ field, we expect $M = N\mu$ for $B \neq 0$. Thus, the very moment that the system is isolated from the surrounding, the magnetic moment should change from $M = N\mu \tanh(\frac{\mu B}{kT})$ to $M = N\mu$. No further change occurs to the magnetic moment, $M = N\mu$, as the $B$ field is reduced towards zero. However, the magnetic moment switches its direction when the $B$ field direction is reversed. Our analysis shows that for an isolated spin system, its magnetic moment is either $N\mu$ or $-N\mu$ controlled completely by the direction of the field. Reversing the direction of the field will reverse the direction of the magnetic moment. If the magnetic moment is locked to $M = N\mu \tanh(\frac{\mu B}{kT})$ as the field is rapidly reversed, the spin system cannot be in an isolated state. In fact, the magnetic moment is not locked, it changes very slowly in comparison to the speed the field changes. In other words, the field changes too rapidly for the system to follow. In thermal systems, all reversible processes must also be quasi-static processes that change slow enough so that the system is always close to equilibrium. The fast switching of the magnetic field applied to the spin system is not a quasi-static process. It is, in fact, an irreversible process that sends the system to a far from equilibrium state. Equations such as Eq. (5) and Eq. (6) are applicable to reversible processes only. If an irreversible process is conducted to the system and Eq. (6) is then applied to find the temperature of the system, the only outcome is an incorrect temperature.

For people not so familiar with the spin system, we will use a more familiar classical system to clarify why one should refrain from applying the standard thermodynamic functions to an irreversible process. Figure 1(a) shows a container with two compartments connected by a narrow tube with radius *r*. Each compartment is a hollow cylinder of radius *R* and height *h*. The two compartments separated by a distance *H* are placed in the Earth's gravitational field, and they are in good thermal contact with a reservoir at a constant temperature *T*. At equilibrium, the number of molecules in the lower and upper compartment are $N_0$ and $N_0 \exp^{-\frac{mgH}{kT}}$, respectively. If we make *h* as small as the ~diameter of the molecules, we can ignore the concentration of molecules as a function of height within either the lower or the upper compartment. Thus, we effectively have a two-level system in terms of the gravitational energy. By rapidly turning the system upside down, we can force the system into a state shown in Fig. 1(b). Now, we expect molecules from the upper compartment to drop down to the lower compartment. If *r* is very small, it will take a relatively long time, a long relaxation time, to establish new equilibrium. At the moment immediately after the system is inverted, we have apparent population inversion. If we apply the Boltzmann distribution, the population inversion would suggest a negative temperature for the system. If we attempt to use Eq. (4) to find the temperature of the system shown in Fig. 1(b), we will also get a negative value. This is easily seen because the addition of *dU* to the system would lift more molecules from the lower to the upper compartment and hence reducing *entropy S*.

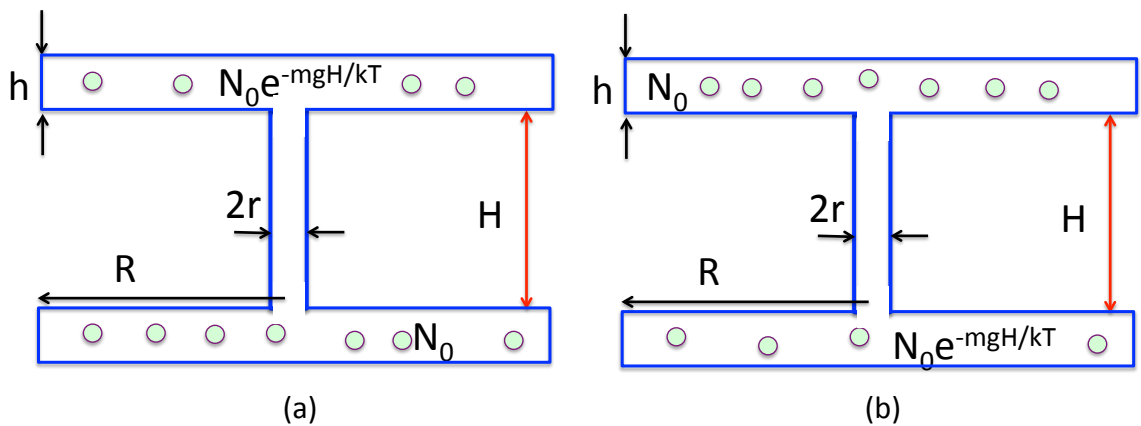

Figure 1. A system of an ideal gas in thermal contact with a reservoir at temperature *T* and in the presence of a gravitational field. The system consists of two identical hollow cylinders connected by a narrow tube of radius *r*. (a). Equilibrium state. Number of molecules in each cylindrical compartment follow the Boltzmann distribution. (b) The same system immediately after being flipped is in a population inverted state.

Ignore the mass of the container, flipping the system upside down leads to $\left(N_0 - N_0 \exp^{-\frac{mgH}{kT}}\right) mgH$ of potential energy increase. Therefore, during flipping, an amount of work equal to $\left(N_0 - N_0 \exp^{-\frac{mgH}{kT}}\right) mgH$ has been done to the system to overcome the gravitational field. If we wait for a long enough time, the stored potential energy will gradually be converted into kinetic energy of the molecules heating the system up. Instead of rapid flipping of the system, we can also try to turn the system upside down in a reversible manner by titling the system very slowly to allow the system to maintain equilibrium. In this case, the system will never enter the population-inverted state. This would also be true for the spin system. If the magnetic field is changed reversibly allowing the system to be close to equilibrium all the time, no population-inverted state should appear. The temperature of the system can be evaluated using Eq. (6) throughout the process as field is reduced and then inverted except for one single point when field is zero.

Another relevant example is the sudden expansion of a gas. Consider a gas in a cylinder with a movable piston. If the piston were suddenly moved so that the volume of the cylinder is increased, the gas would expand towards the new volume created. Imagine that the piston is moved much more quickly than the gas is able to follow. The system would momentarily enter a state where the gas occupies the initial volume and the newly created volume being vacuum. If we insist to use $p = T\left(\frac{\partial S}{\partial V}\right)_{U,N}$ to evaluate the pressure of the system, we would get zero pressure, which is not correct.

A thermal system would respond to changes to its volume, energy or magnetic field for a spin system. The response has an intrinsic relaxation time. If changes were made much faster than the relaxation time, certain properties of the system become not well defined as the system is undergoing transition. For the population inverted spin system, its magnetic energy is well defined, but its temperature is not. For the rapid volume change of an ideal gas, its energy is well defined, but its pressure is not. The thermal dynamic functions describing systems at equilibrium should not be applied to systems that are put into a far from equilibrium state via an irreversible process. The approach of Struchtrup [5] is useful in a way it treats "un-useable energy" as stored energy. Energy in a system can be trapped, momentarily or permanently, and unable to participate energy exchange for many different reasons. For the population inverted spin system, the long relaxation time is equivalent to the situation that the magnetic energy cannot be freely released due to some constraint. When applying Eq. (6) to a system, it is assumed that there is no such constraint to energy release.

To summarise, our analysis demonstrates that negative temperature of a system is a direct consequence of incorrectly applying thermodynamic functions to an irreversible process. If the functions were applied to the system at or very close to equilibriums, no negative temperature would ever appear. When considering the population inverted spin system in equilibrium with a black body radiation background, one would require the background has energy density higher than infinity. This makes the negative temperature states physically unrealistic.

**References**
[1] E. M. Purcell, V. Pound, *Phys. Rev.* (1951) **81**, 279.
[2] N. F. Ramsey, *Phys. Rev.* (1956) **103**, 20.
[3] P. Hakonen, O. V. Lounasmaa, *Science*, (1994) **265**, 1821.
[4] Braun et al, *Science*, (2013) **339**, 52-55.
[5] H. Struchtrup, *Phys. Rev. Lett.* (2018) **120**, 250602.
[6] V. Romero-Rochin, *Phys. Rev. E* (2013) **88**, 022144.
[7] J. Dunkel, S. Hilbert, *Nature Phys.* (2014) **10**, 67.
[8] R. H. Sweddsen, *Rep. Prog. Phys.* (2018) **81**, 072001.
[9] E. Abraham, O. Penrose, *Phys. Rev. E* (2017) **95**, 012125.
[10] D. Frenkel, P. B. Warren, *Am. J. Phys.* (2015) **83**, 163.
[11] S. Calabrese, A. Porporato, *Phys. Lett. A* (2019) **383**, 2153.